\documentclass[a4paper,11pt]{article}
\pdfoutput=1 

\usepackage{jinstpub} 

\title{SND electromagnetic calorimeter time measurement and its applications}
\author[a,b]{M.N. Achasov,}
\author[a]{A.G. Bogdanchikov,}
\author[a,b]{V.P. Druzhinin,}
\author[a,b]{A.A. Korol,}
\author[a]{D.P. Kovrizhin,}
\author[a,1]{N.A. Melnikova,\note{Corresponding author.}}
\author[a,b]{S.I. Serednyakov,}
\author[a]{I.K. Surin,}
\author[a]{A.I. Tekut'ev,}
\author[a,b]{Yu.V. Usov,}
\author[a,b]{and V.V. Zhulanov} 

\affiliation[a]{Budker Institute of Nuclear Physics,\\11, Acad. Lavrentieva Pr., Novosibirsk, 630090 Russian Federation}
\affiliation[b]{Novosibirsk State University,\\1, Pirogova str., Novosibirsk, 630090, Russian Federation}

\emailAdd{N.A.Melnikova@inp.nsk.su}

\abstract{%
 The SND is a non-magnetic detector deployed at the VEPP-2000 e+e- collider (BINP, Novosibirsk) for hadronic cross-section measurements in the center of mass energy region below 2 GeV. The important part of the detector is a three-layer hodoscopic electromagnetic calorimeter (EMC) based on NaI(Tl) counters. Until the recent EMC spectrometric channel upgrade, only the energy deposition measurement in counters was possible. A new EMC signal shaping and digitizing electronics based on FADC allows us to obtain also the event time structure. The new electronics and supporting software, including digital signal processing algorithms, are used for data taking in the ongoing experiment. We discuss the amplitude and time extraction algorithms, the new system performance on experimental events and physical analysis applications.
}
\keywords{Calorimeter methods, Calorimeters, Electronic detector readout concepts (solid-state), Instrumentation and methods for time-of-flight (TOF) spectroscopy, Detector modelling and simulations II, Solid state detectors, Timing detectors, Digital signal processing (DSP)}

\begin{document}
\maketitle
\flushbottom
\section{Introduction}
\label{intro}
\begin{figure*}
	\centering
	\includegraphics[width=0.45\columnwidth]{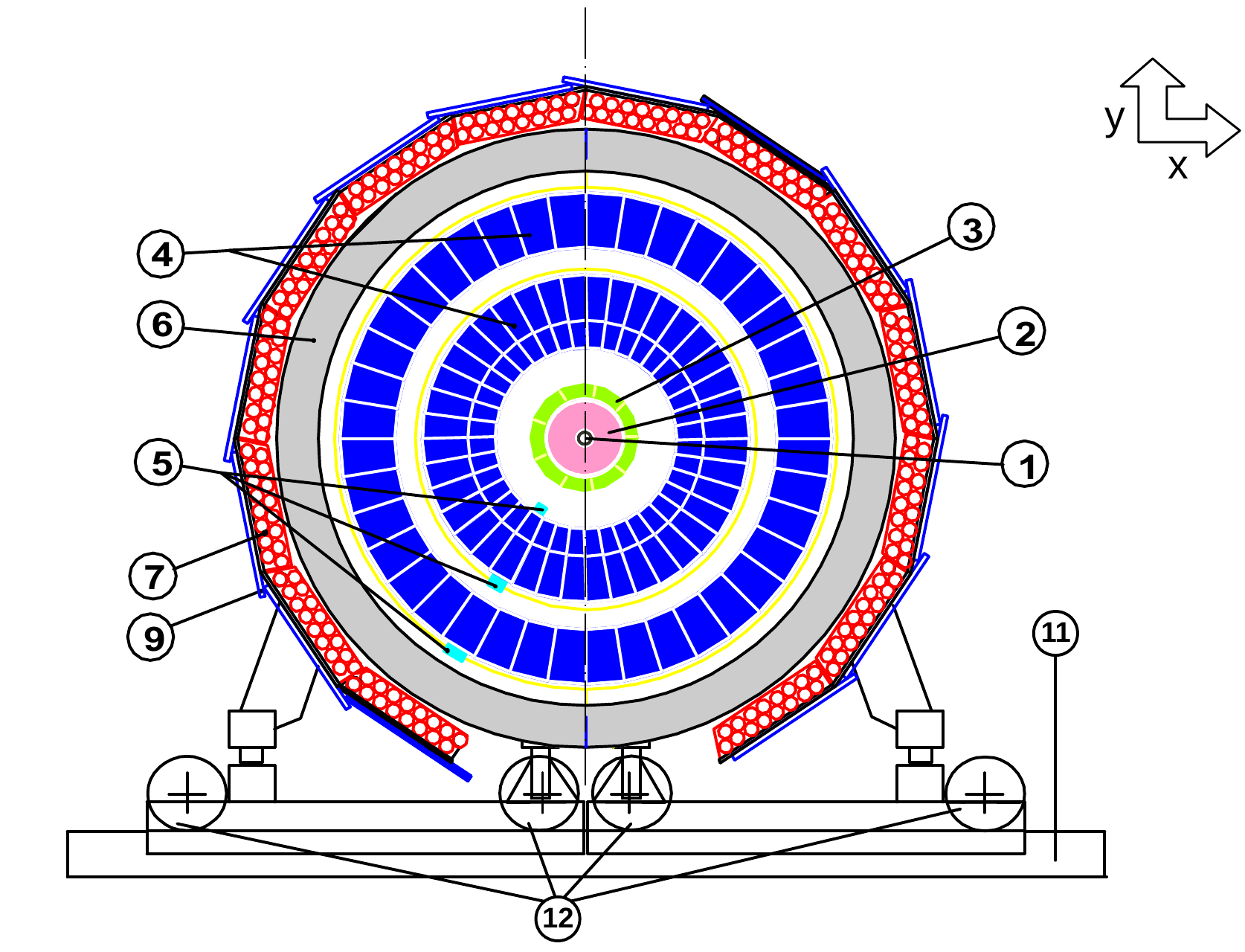}
	\includegraphics[width=0.45\columnwidth]{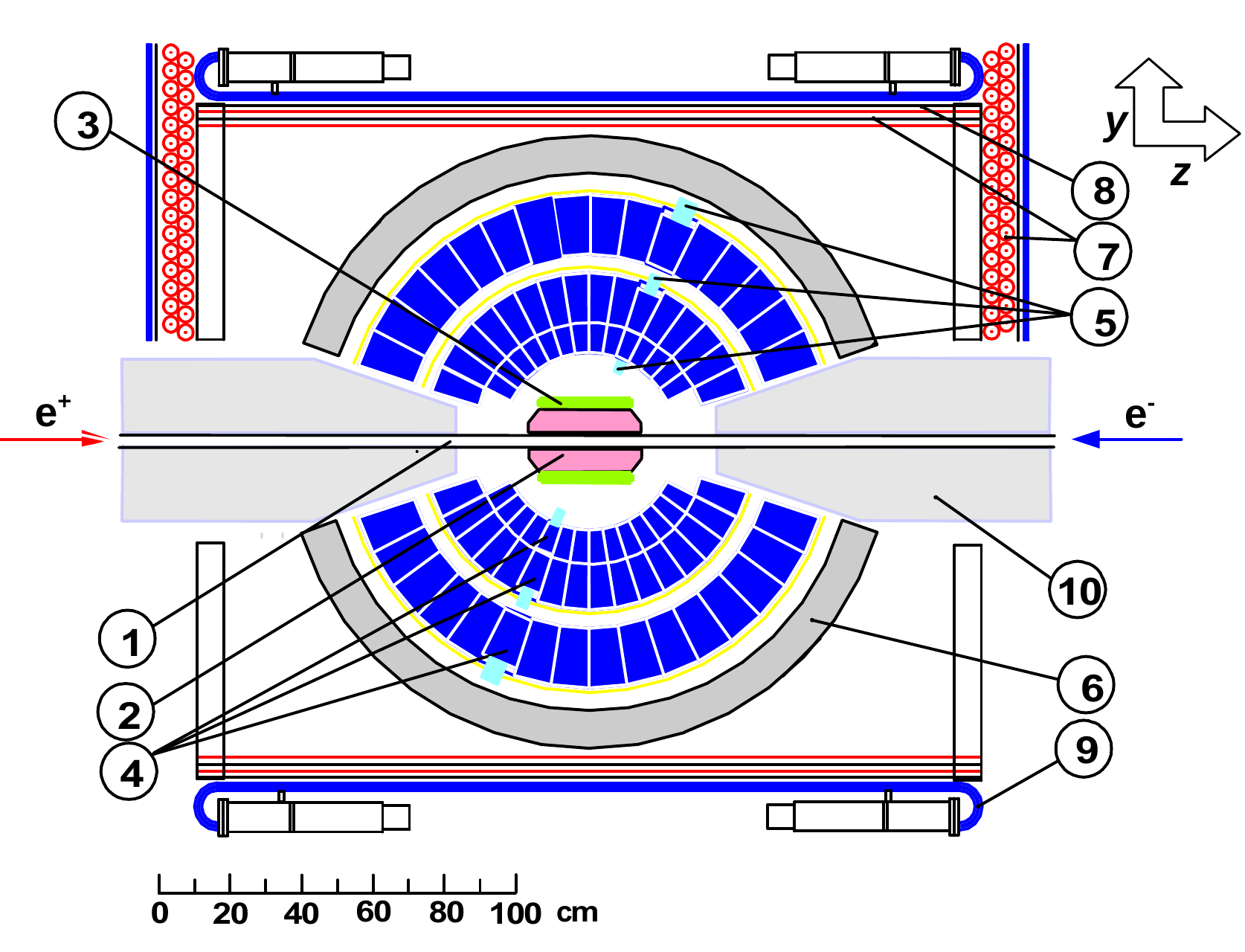}
	\protect\caption{The SND scheme: 1 \textemdash{} vacuum pipe, 2 \textemdash{} tracking
		system (TS), 3 \textemdash{} threshold Cherenkov counter, 4\textendash 5
		\textemdash{} electromagnetic calorimeter (NaI (Tl)) (EMC), 6 \textemdash{}
		iron absorber, 7\textendash 9 \textemdash{} muon detector, 10 \textemdash{}
		focusing solenoids, 11 \textendash{} rails, 12 \textendash{} wheels.
		\label{fig:SNDScheme:}}
\end{figure*}

The Spherical Neutral Detector (SND) \cite{snd1,snd2,snd3}  is a general purpose nonmagnetic detector employed for hadronic cross section measurement experiments, studies of hadron production dynamics and etc at the $e^{+} e^{-}$ collider VEPP-2000 \cite{vepp2k}. The collider operates in the center of mass energy range from $0.3$ to $2$ GeV using round beam optics. 
The SND has a typical layout (figure~\ref{fig:SNDScheme:}) with a cylindrical tracking system, an electromagnetic calorimeter (EMC), threshold Cherenkov counters and a muon detector. 

The EMC is the most important and massive part of the detector, it provides uniform particle detection in a solid angle of $0.95 \cdot 4\pi$. The EMC consists of three layers of $1632$ counters, each counter includes a NaI(Tl) crystal and its vacuum phototriod (VPT). EMC counters form $160$ towers, where one tower consists of $12$ neighboring  counters from all three EMC layers.

The EMC spectrometric channel has been upgraded recently. The new electronics allows us to obtain a digitized signal waveforms which are processed to extract energy and time information.

\section{The EMC spectrometric channel}
\label{sec-1}
\subsection{Upgraded EMC electronics}
\label{subsec-1-1}
The schematic layout of the new  EMC spectrometric channel \cite{newChannel_nim2016} is shown in figure~\ref{fig:emc_channel:}. It consists of a EMC counter, a charge-sensitive preamplifier (CSA), a shaper (F12 module) and a digitizer (Z24 module). 

Each F12 module shapes signals from one EMC tower. It also provides four fast  signals for a first level trigger (FLT) system:  tower total energy deposition signal (analog) and signals of total energy deposition in each tower layer after discriminators. These fast signals from all F12 modules are combined to produce an analog EMC total energy deposition signal and EMC logical signals which are used to form the FLT signal responsible for the start of the read-out process.

The shaped signals are digitized by Flash-ADCs (FADCs) inside new Z24 modules.
The Z24 module includes a Xilinx system on chip and six FADCs with four $12$-bit channels each. The  digitized waveforms are read out after the arrival of a first-level trigger (FLT) signal. The FLT signal is synchronized with a beam revolution frequency ($f_\mathrm{br} = 12.3$ MHz). The signal sampling is performed at the frequency of $3\,f_\mathrm{br}$. 
The Z24 module represents the main difference between the new and the old EMC channel that provided only peak amplitude measurements. The new EMC electronics has been in use for the data taking since the end of 2018 year, making possible time measurements with the SND calorimeter. For now, all digitized signal waveforms are processed on an online computing farm and also stored for further offline re-processing.

\subsection{EMC signal properties}
\label{subsec-1-2}
\begin{figure}[h]
	\centering
	\includegraphics[width=0.99\columnwidth]{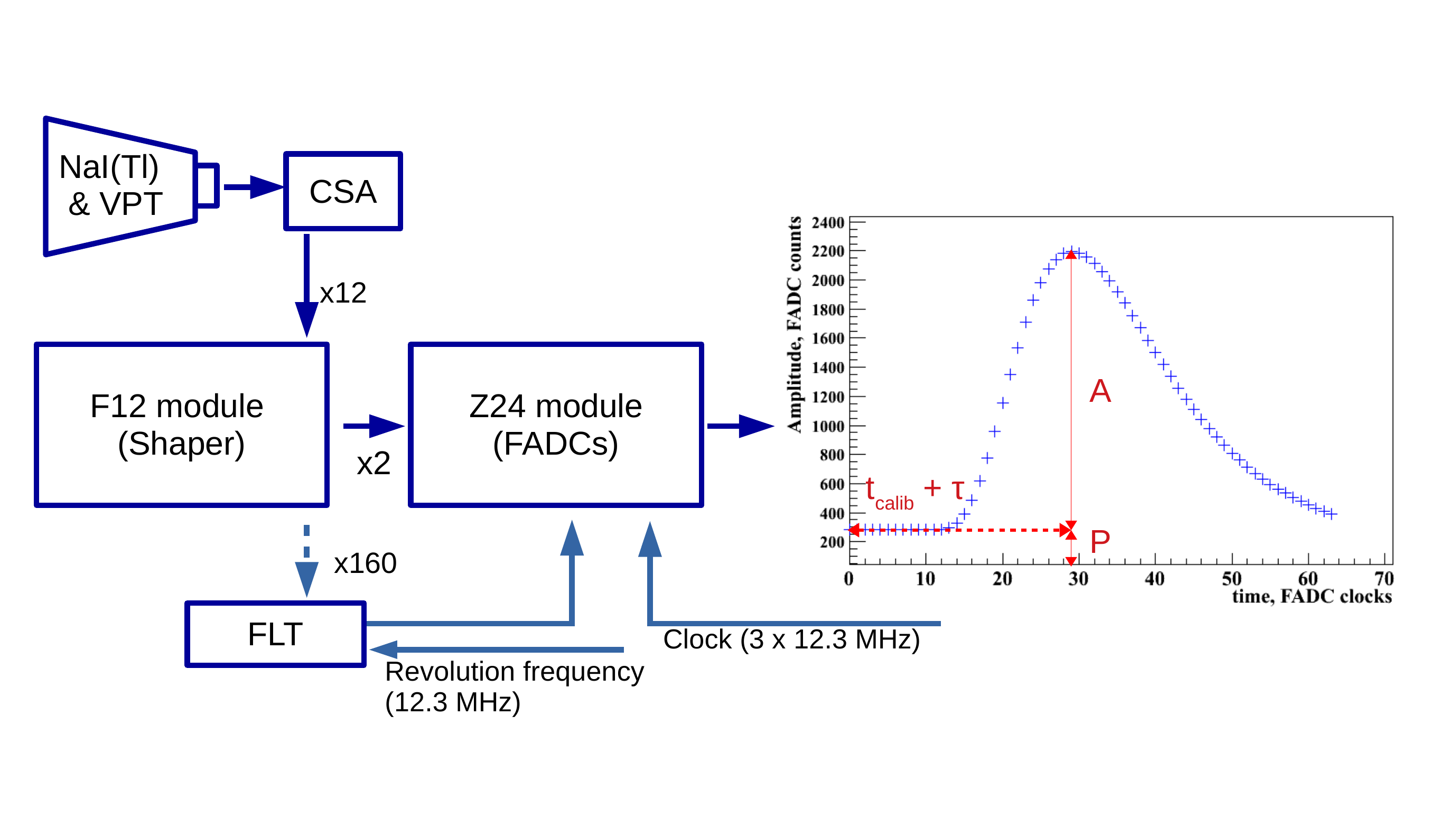}
	\protect\caption{Schematic layout of the EMC spectrometric channel  and a typical EMC signal shape. The abbreviations are described in the text.}
	\label{fig:emc_channel:}
\end{figure}

A typical digitized EMC signal waveform with $64$ samples and the sampling period ($ T_\mathrm{s}$) of  $\approx 27$ ns is shown in figure~\ref{fig:emc_channel:}. Typical pedestal values are $ \sim300$ FADC counts with pedestal noise of $3\div5$ counts (one FADC count is equal to $\sim0.25$ MeV depending on the EMC channel). 

F12 modules provide stable pulse shaping for all counters, but the signal waveform may differ from one EMC channel to another. A dedicated calibration procedure is used to retrieve signal waveforms for all EMC channels (section~\ref{sec-4}).
We describe the EMC signal waveform using a function $U(t)$:
\begin{equation}\label{sigFunc}
U(t) = A \cdot F(t - \tau) + P
\end{equation} 
where $A$ \,---\, the signal amplitude, $F(t)$ \,---\, the function of calibrated signal shape in the corresponding channel, $\tau$ \,---\, the signal arrival time with respect to the function $F$ and $P$ \,---\, the signal pedestal. 

\section{EMC time simulation and reconstruction}
\label{sec-2}
\subsection{EMC channel response simulation}
\label{subsec-2-1}
EMC simulation software has been modified to imitate realistic response of the new electronics. 
We use the Geant4 framework \cite{Geant4} to simulate particle passage through matter. Each EMC crystal is divided into several sensitive cells. As a result, several cell hits are generated for each triggered crystal. Available information about  all cell hits in the crystal is used to calculate the response signal magnitude $U_{i}$ at each sampling point:
\begin{equation}\label{simFuncc}
U_{i} = k_\mathrm{chan} \cdot \sum_{j}{ (E_j \cdot F(i \,-\, t_j \,-\, t_\mathrm{shift}))\,+\, P \,+\, P_\mathrm{noise}}
\end{equation} 
where $k_\mathrm{chan}$ \,---\, the conversion coefficient from FADC counts to MeV in the channel, $E_j, t_j$ \,---\, energy deposition and the hit time of the $j$ cell hit correspondingly, $t_\mathrm{shift}$ \,---\, the common time shift that can be set, $P$ \,---\, the channel pedestal calibrated value, $P_\mathrm{noise}$ \,---\, the channel pedestal noise calibrated value (the same for all samples). The hit time is calculated relatively to the event start time using Geant4 core pre-step and post-step time points ($t_\mathrm{prestep}$, $t_\mathrm{poststep}$) and (if necessary) the light time of flight from the cell center to the phototriode plane ($t_\mathrm{flight}$): $t_j = (t_\mathrm{prestep} \:+\:  t_\mathrm{poststep}) \: / \: 2 \: + \: t_\mathrm{flight}$. For $t_\mathrm{flight}$ calculation we consider that light propagates in straight-lines along the radius, this model is far from exact but allows us to simulate minimal time values. 

\subsection{EMC time event reconstruction}
\label{subsec-2-2}
Reconstruction with EMC time has been implemented in the SND offline framework in a simple way. At first the usual reconstruction is performed. Then the results of an EMC clusterization algorithm is used to obtain the list of the EMC hits that belongs to each particle EMC cluster. Using time information of these hits, an EMC time object is constructed for each particle candidate. As a result new entities can be used in analysis: EMC counter time, EMC particle time, EMC event time.

\section{EMC signal processing}
\label{sec-3}
The values of the signal parameters ($A, P, \tau$) are extracted  by fitting waveforms to the function $U(t)$. 
At this moment we have two algorithms for the fitting. The first one is based on the linearization and is used for fast processing most of the signals. The second one is an alternative algorithm based on the correlation function that is used to handle special cases that can't be processed with the first solution.

\subsection{The linearization algorithm}
\label{subsec-3-1}
The linearization algorithm \cite{LinAlgo2017} is an adapted version of the algorithm developed for the Belle II calorimeter electronics \cite{Belle2algo}. Its design ensures that it can be applied inside the Z24 module on FPGA. The algorithm minimizes the $ \chi^{2} $ function with floating parameters $A$, $P$ and $\tau$:
\begin{equation}\label{algo1}
\chi ^{2}(t) \,=\, (y_{i} \,-\, A \cdot F(t_{i} \,-\, \tau) \,-\, P) \cdot S_{ij}^{-1} \cdot (y_{i} \,-\, A \cdot F(t_{j} \,-\, \tau) \,-\, P)
\end{equation} 
where $y_{i}$ \,---\, the signal sample magnitude at $t_{i}$ time point with $i \in [0, 63]$, $F(t_{i})$ \,---\, the calibrated signal shape function (cubic B-spline), $S$ \,---\, the noise covariance matrix. The function $F(t)$ is linearized on a time grid with time step of  $\frac{1}{50} \, T_{s} $, allowing all needed for minimization coefficients to be calculated in advance. The initial value for $\tau$ is estimated using the maximum sample time point of the signal waveform. The algorithm converges usually after $1-2$ iterations with maximum number of allowed iterations being set to $3$. The chosen time grid step allows us to achieve time resolution required for physical analysis (see section~\ref{sec-5}) and to keep computational costs at the reasonable level.

This algorithm is relatively fast ($\sim0.15$ ms per signal) and reliably processes most of the signals with estimated $50 < A < 4095$ FADC counts and $\tau$ in the range $[-7, 7]$ FADC clocks ($\sim8.5\%$ of all hits).

Only values for $A$ and $P$ are obtained for the signals of estimated $A < 50 $ FADC counts using the linear regression model with $\tau$  being fixed. Most of the fittings with fixed time are successful ($\sim60\%$ of all hits), for signals with very small $A$ or shifted peaks the processing results in zero or negative amplitude values ($\sim30\%$ of all hits). 

Strong signals with heavily shifted peaks ($> 7$ FADC clocks) or with FADC saturation ($\sim0.1\%$ of all hits) and other shape distortions can't be processed correctly using the described algorithm without its complication. 

\begin{figure*}
	\centering
	\includegraphics[width=1.\columnwidth]{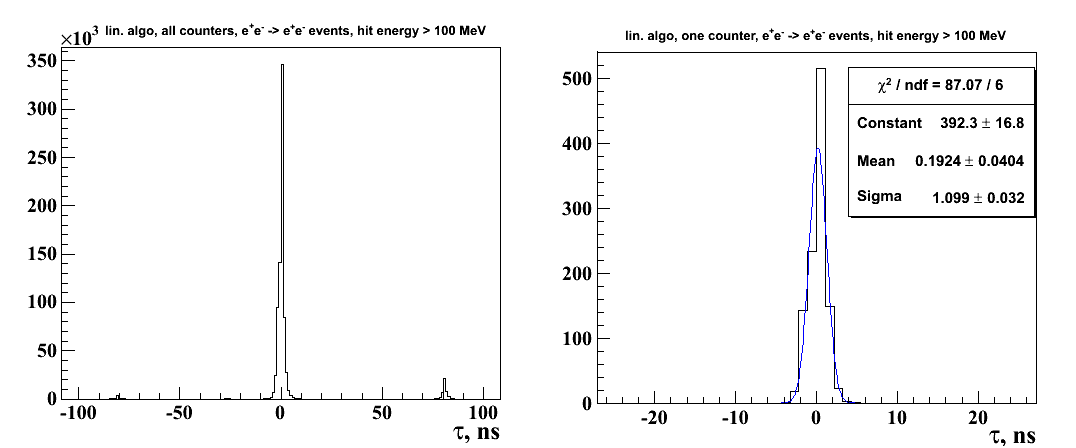}
	\protect\caption{The linearization algorithm results on $e^{+}e^{-} \to e^{+}e^{-}$ events. On the left: the signal times for all triggered EMC counters with the main peak around zero as it should be due to the signal shape calibration procedure being performed on the same process. On the right: the obtained time resolution in one EMC counter on fitted signals with reconstructed energy deposition $E > 100$ MeV.   
		\label{fig:linalgores:}}
\end{figure*}

The result of this algorithm on data is shown in figure~\ref{fig:linalgores:}. There are several peaks can be seen in the  time spectrum obtained on $e^{+}e^{-} \to e^{+}e^{-}$ events. The main peak is near zero because we calibrate EMC signal waveforms in all EMC channels using this type of events. The peak near $81$ ns represents beam-induced background, which should be expected at $n\cdot 3 \cdot T_\mathrm{s} (n \in \mathbb{Z})$ time points. 

\subsection{The correlation function algorithm}
\label{subsec-3-2}
The correlation function algorithm processes a signal in two steps. 
At the first step it finds the maximum of the correlation function  ($\omega(t)$) between the signal ($y_{i}$) and the calibrated signal waveform (cubic B-spline $F$) in the corresponding EMC channel to determine the time shift ($\tau$) between them:
\begin{equation}\label{corfunc}
\omega(\tau) = \sum_{i=0}^{i=63} (y_{i} \,-\, P) \cdot F(t_{i} \,-\, \tau).
\end{equation} 
On practice, we minimize $-\omega(t)$  using the GNU Scientific Library (GSL) \cite{GSL_manual} implementation of the Brent's algorithm and GSL fast Fourier fransform (FFT) methods for faster calculations. The first guess for the time shift ($\tau_0$) is calculated using the discrete Fourier transform shift property:
\begin{equation}\label{fisrtGuess}
\widehat{y_i-P}(\omega) = exp(-i\omega\tau_0)\cdot \widehat{F_i}(\omega),\quad  \omega = \frac{2\pi}{64} 
\end{equation} 
where the hat denotes the output of the Fourier transform.
At the second step, the obtained $\tau$ value is applied to shift $F(t)$ along the time axis, and the values of $A$ and $P$ are extracted using the linear regression model.

\begin{figure*}
	\centering
	\includegraphics[width=0.49\columnwidth]{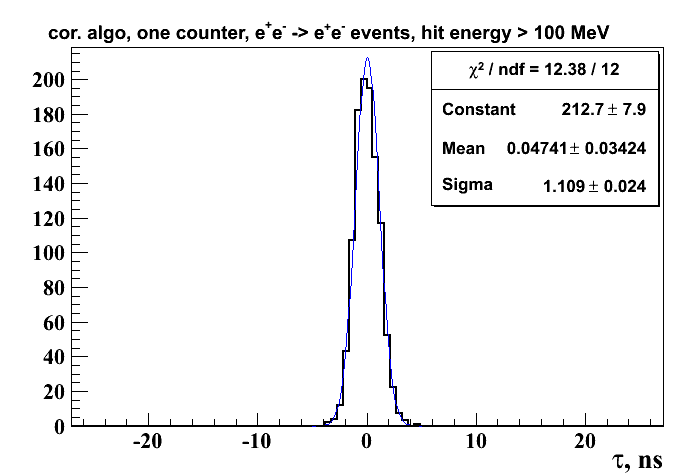}
	\includegraphics[width=0.49\columnwidth]{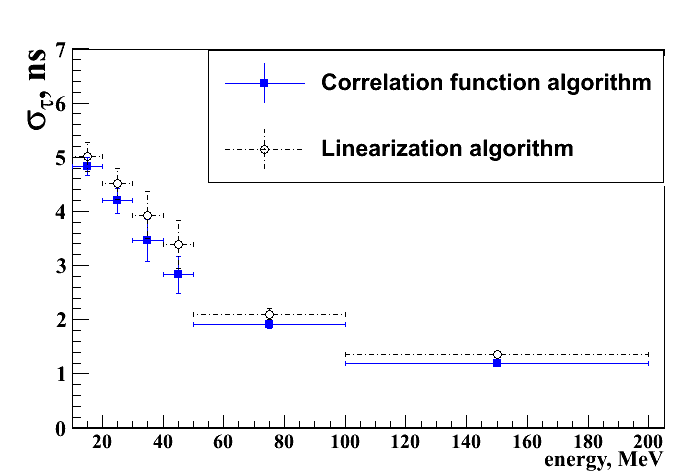}
	\protect\caption{The correlation function algorithm results on $e^{+}e^{-} \to e^{+}e^{-}$ events. On the left: the obtained time resolution in one EMC counter on fitted signals with reconstructed energy deposition $E > 100$ MeV for the main time peak. On the right: time resolution dependence on EMC energy deposition for the same EMC counter obtained by two algorithms.   
		\label{fig:corfuncres:}}
\end{figure*}

The obtained time resolution for one EMC counter and its dependence on EMC energy deposition is almost the same as we achieved with the linearization algorithm ( figure~\ref{fig:corfuncres:}). This algorithm can successfully process almost all signals, but it's relatively  slow ($\sim1.2$ ms per signal) and gives bad time resolution for small amplitudes. 

At this moment we apply the correlation function algorithm for processing heavily shifted ($> 7$ FADC clocks) and saturated signals. Shifted signals can occur as a result of previous event pile-ups (negative time values) or from nuclear interactions with matter (positive time values). The algorithm was tuned for these cases and its performance has been validated on MC signals with known properties. 

\begin{figure*}
	\centering
	\includegraphics[width=0.4\columnwidth]{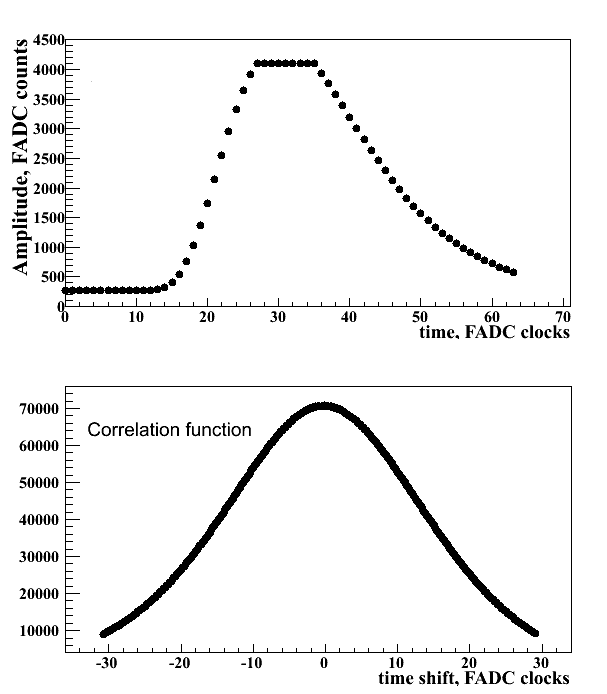}
	\includegraphics[width=0.49\columnwidth]{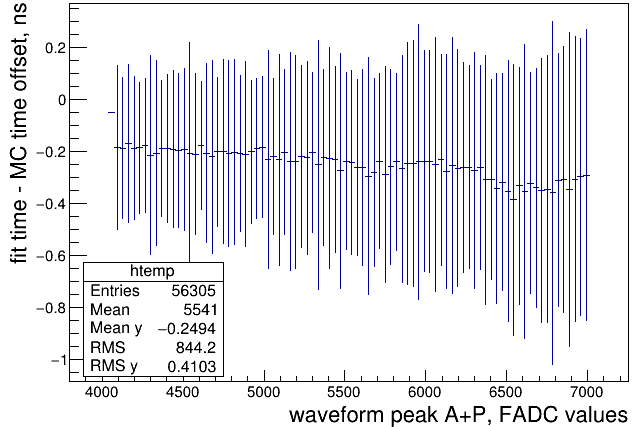}
	\protect\caption{The correlation function algorithm on saturated waveforms. On the left: a typical EMC saturated signal waveform and its correlation function. On the right: the spread profile for $\tau \,-\, t_\mathrm{shift}$ versus MC waveform magnitude ($A\,+\,P$) before saturation cut off, here the same time offset ($t_\mathrm{shift}$) was set to shift all waveforms. 
		\label{fig:saturperforms:}}
\end{figure*}

To process saturated signals we imitate the saturation of the calibrated waveform ($F(t)$) for calculating FFT first guess ($\tau_0$). At the second step,  only  the signal amplitude is determined with $P$ being fixed to the calibrated value and the saturated signal samples being ignored.
The algorithm performance on the saturated MC signals is shown in figure~\ref{fig:saturperforms:}.

\begin{figure*}
	\centering
	\includegraphics[width=0.4\columnwidth]{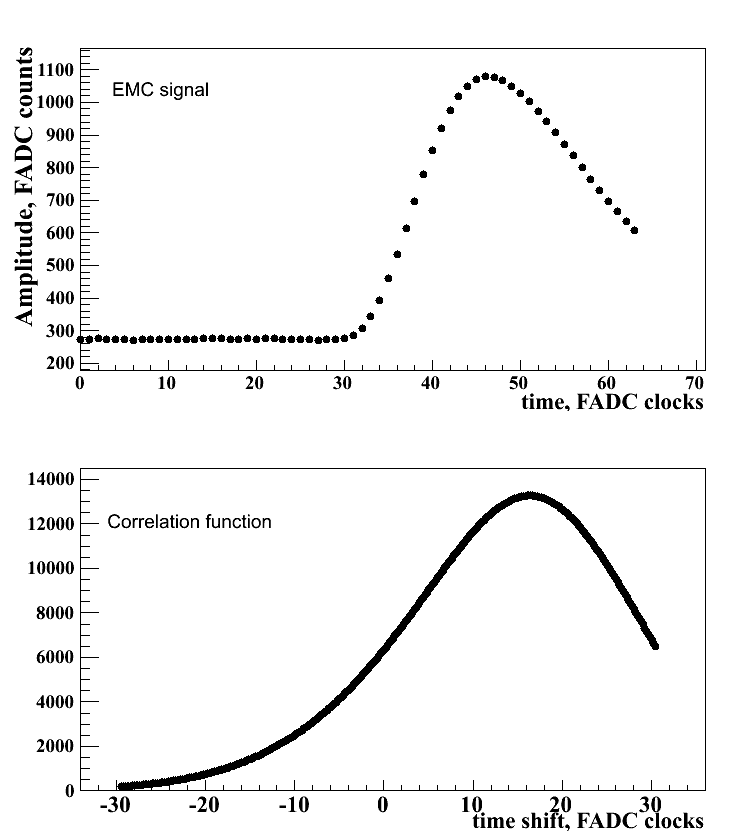}
	\includegraphics[width=0.49\columnwidth]{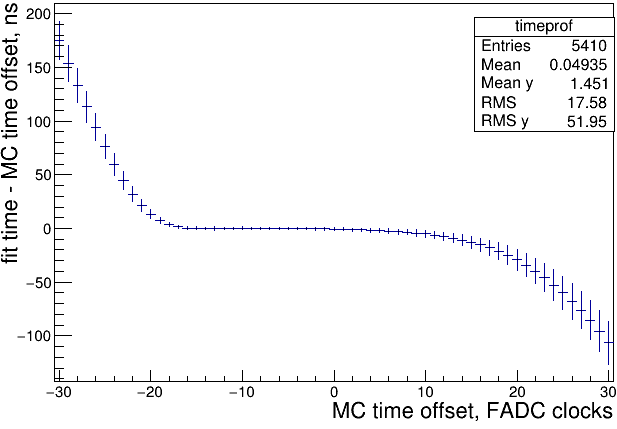}
	\protect\caption{The correlation function algorithm on shifted waveforms. On the left: a real shifted EMC signal waveform and its correlation function. On the right: the spread profile for $\tau \,-\, t_\mathrm{shift}$ versus used MC  $t_\mathrm{shift}$ that was set in range $[-30, 30]$  FADC clocks. 
		\label{fig:shiftedperforms:}}
\end{figure*}

In case of heavily shifted signals only $A$ is calculated at the algorithm second step with $P$ fixed to the calibrated value. The algorithm performance on MC signal waveforms simulated with the set time shift $t_\mathrm{shift} \in [-30, 30]$ FADC clocks in shown in figure~\ref{fig:shiftedperforms:}.

\section{EMC signal waveform calibration}
\label{sec-4}
As it was mentioned in subsection~\ref{subsec-1-2} the key moment of EMC time measurements is stability of reference signal waveforms in all EMC channels that we calibrate using special procedures \cite{shapeCalib2019}. For calibration three types of data are used: calibration generator pulses, cosmic muons and $e^{+}e^{-} \to e^{+}e^{-}$ events. 

To perform the calibration procedure with the EMC generator one hundred signal pulses with known peak locations and $A$ values are generated for each EMC channel. These signals are used to construct an averaged normalized pulse that is fitted with a cubic B-spline. The obtained B-spline coefficients are stored in the calibration database and can be used later to restore a reference signal waveform. This procedure is very fast and performed daily during data taking. Unfortunately, the result differs significantly from real event signal waveforms and can't be applied to process them accurately so it's used mostly for monitoring purposes.

To process real signals the result of the waveform calibration procedure on $e^{+}e^{-} \to e^{+}e^{-}$ events is used. This is a very difficult and sensitive to the input data process. It's performed on large data sets, with special efforts being made to carefully select only strong well-shaped signals from the main event time peak ($t_{event}$). The procedure iterates over 2 steps. At the first one, the cubic B-spline of an initial reference waveform is used to fit  the selected signals and determine values of $A$ and $\tau$. At the next step, signals with $A > 20$ MeV and $\tau$ close to $t_{event}$ are used to construct an averaged normalized pulse with zero pedestal value. This pulse is fitted by a cubic $16$ basis B-spline (GSL implementation). The obtained B-spline coefficients are stored in the SND database for signal processing or used to reconstruct the reference waveform cubic B-spline for the next iteration. 

For the very first iteration the previously obtained calibration results are used. In the absence of those, the initial waveforms are obtained by performing the calibration on cosmic muons. It's the same procedure in most ways, except input data selection. Since time in these events is uniformly distributed, the procedure selects the most strong and well-shaped signal pulse as a reference one to obtain the initial waveform. The waveform calibration on cosmic muons is performed after long stops in operation or maintenance work.

\section{Time measurement applications}
\label{sec-5}

EMC time measurement gives us the opportunity of identifying interesting events using the time-of-light technique. EMC time can be used for cosmic-ray background suppression due to uniform time distribution of these events. Beam-induced background can be suppressed as it mostly located  at $n \cdot 3 \cdot T_\mathrm{s}$ $(n \in \mathbb{Z})$ time points (section~\ref{subsec-2-1}, figure~\ref{fig:linalgores:}). 
Time filters can be applied to identificate  events with long EMC response times e.g. $e^{+}e^{-} \to n\bar{n}$ \cite{nan_art2009, nan2019}. Neutron-antineutron pairs in the energy range of VEPP-2000 are non-relativistic so it could take about $8$ ns for an antineutron with kinetic energy of $5$ MeV to annihilate in the first EMC layer.

\begin{figure*}
	\centering
	\includegraphics[width=0.7\columnwidth]{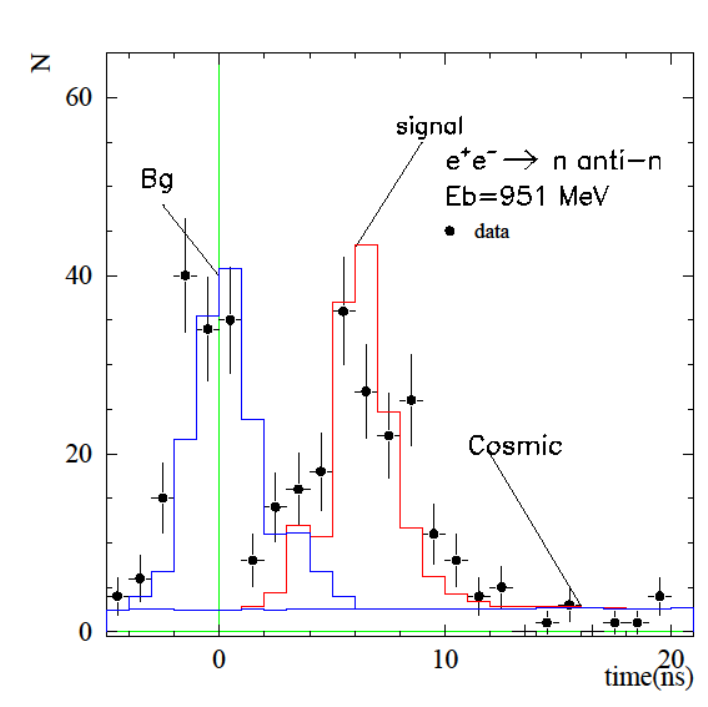}
	\protect\caption{EMC time spectra for $e^{+}e^{-} \to n\bar{n}$ (signal), background events ($e^{+}e^{-} \to p\bar{p}$ and $e^{+}e^{-} \to \gamma\gamma$)  and cosmic events. Markers with bars represent data, histograms \,---\, MC data. 
		\label{fig:nan_pics:}}
\end{figure*}

The first attempt was made to identify these events using EMC time. Preliminary results are shown in figure~\ref{fig:nan_pics:}, that demonstrates good separation between $e^{+}e^{-} \to n\bar{n}$ events and background events ($e^{+}e^{-} \to p\bar{p}, \gamma\gamma$) near the threshold.

\section{Summary}
The new EMC electronics provides digitized signal waveforms that are processed by two dedicated algorithms to extract amplitude and time information: the linearization algorithm and the correlation function algorithm. The linearization algorithm is faster and successfully processes most of the signals while the correlation function algorithm is slower and used to handle shifted and saturated signals that can't be processed with the linearization algorithm without its complication. The algorithm performance for these two cases was validated on MC data with new EMC electronics response simulation. The performance of the algorithms on data recorded during the 2018--2019 experimental season is presented. The achieved time resolution in the EMC counter is $\sim 1$ ns for energy deposition of $100$ MeV. The main physical analysis applications are discussed. The preliminary results for $e^{+}e^{-} \to n\bar{n}$ event identification using EMC time are demonstrated as an example. 

\section{Acknowledgments}
The work is partially supported by the Russian Foundation
for Basic Research (project nos. 18-02-00382-a, 18-02-00147-a and 20-02-00347).

\bibliographystyle{unsrt}
\bibliography{article_bib}

\end{document}